# AC-Augmented Dielectric Barrier Discharge

Anthony Tang[1], Alexander Mamishev[2], Igor Novosselov[1,3,*]

*[1]Department of Mechanical Engineering, University of Washington, Seattle, U.S.A. 98195*

*[2]Department of Electrical and Computer Engineering, University of Washington, Seattle, U.S.A. 98195*

*[3]Institute for Nano-Engineered Systems, University of Washington, Seattle, U.S.A. 98195*

## ABSTRACT

Dielectric barrier discharge (DBD) plasma actuators generate an electrohydrodynamic (EHD) force through the ionization and acceleration of charged species. Most active flow control DBD applications are only practical at lower Reynolds numbers, and increasing the momentum injection can extend the practical uses of the technology. Here, we experimentally demonstrate improvement in the performance of a planar DBD actuator by utilizing an AC-augmented electrical field in a three-electrode geometry. Time-resolved electrical and optical measurements, velocity profiles, and direct thrust measurements were used to characterize the EHD augmentation. Varying phase shift and E-field strength between the two air-exposed DBD electrodes can accelerate EHD flow and increase EHD forcing by up to ~ 40%. At the most favorable conditions, the maximum thrust was 54 mN/m when the air-exposed electrodes were out of phase. In-phase operation of the exposed electrodes at high E-field conditions can induce adverse effects and sliding discharge. Mechanistically, the performance improvements in the AC-augmented DBD actuator primarily come from the additional charge pull action by the third electrode. The insight into the AC-augmented DBD mechanism allows for developing multi-stage arrays capable of further increasing EHD forces.

[*]corresponding author ivn@uw.edu

# 1. Introduction

Non-thermal plasma devices have received significant interest in scientific and engineering applications [1, 2]. Plasma discharge has been used in surface disinfection [3, 4], particle charging for two-stage electrostatic precipitators [5, 6], and ionization sources for mass spectrometry applications [7]. Plasma devices can also generate fluid momentum that could be utilized for propulsion [8-11] and flow control [12, 13]. Non-thermal corona discharge and dielectric barrier discharge (DBD) generate a flow by ionizing gas molecules and accelerating the charged species in the electric field. This electrohydrodynamic (EHD) action is generated through collisions with neutral molecules. In corona discharge, the ionization happens between two high voltage (HV) direct current (DC) electrodes [8, 14, 15]. In contrast, DBD actuators use HV pulsed DC [16] or alternating current (AC) [17].

A conventional single-stage DBD actuator consists of an electrode pair: an air-exposed electrode and an embedded electrode, with a dielectric barrier between the two. When an AC voltage is applied between the electrodes to generate an electric field greater than the breakdown strength of air, the molecules surrounding the air-exposed (active) electrode become ionized and are accelerated downstream. The momentum injection by the DBD actuators has been characterized by several investigators, e.g., [18-20]. However, the time-dependent multiphysics interactions responsible for energy transfer in electromechanical systems such as DBD are poorly understood due to the complex coupling between electric field, surface and space charging, species recombination, and charge–flow interaction. The plasma–flow interaction can be divided into positive- and negative-going voltage half-cycles. One of the most accepted forcing mechanisms is the 'push-push' mechanism: positive ions are repelled from the active electrode in the positive half-cycle, and negative species (predominantly electrons) are repelled in the negative half-cycle [21]. In reality, these interactions are more complicated. Current measurements and high-speed visualizations show that the positive half-cycle produces streamer-like discharge, and the negative half-cycle shows more frequent smaller peaks associated with glow discharge [22, 23]. The discharge current is typically higher in the positive half-cycle [18]; however, some reports suggest that the forcing terms can be as significant in the negative half-cycle due to the high electron mobility [19, 24].

Despite their relatively low electromechanical efficiency, researchers proposed the DBD actuators for use in aerodynamic applications, such as drag reduction [25, 26], lift augmentation [27, 28], separation control [29, 30], and electric propulsion [9, 10, 31, 32]. Due to their relatively weak momentum injection, the DBD applications are limited to low Reynolds (Re) numbers [33, 34]. Most experimental reports focus on single two-electrode DBD actuators with varied electrode spacing, driving waveforms, voltages, and frequencies. Optimization of physical parameters such as dielectric thickness and active electrode shapes can improve the thrust up to ~50%. For example, Thomas *et al.* [35] reported increased thrust by using thick dielectrics (> 5 mm) due to the reduction in streamer formation. Before streamer formation, thrust scales with applied voltage approximately as the second-order polynomial [18, 36, 37]. Electrical parameters such as waveforms and frequencies have been explored for traditional two-electrode DBD actuators [38]. In a standard two-electrode DBD, symmetric waveforms (square and sine waves) typically produce the highest thrust, and a sine waveform is more efficient than a square waveform [21, 39]. An optimal frequency range for DBD was reported to be 1.5 - 4 kHz [21]. While higher frequencies produce greater total discharges, the electromechanical efficiency diminishes due to a higher rate of ion recombination and a limited acceleration time [18].

DBD forcing on a fluid volume can be increased by introducing a biased third electrode [40, 41]; both positive and negative biases were previously reported [42, 43]. In this arrangement, the DBD electrode pair ionizes the gas, and the third electrode accelerates positive or negative species, promoting their interaction with neutral molecules. Tang et al. studied the plasma/ flow interaction in a DC-augmented (DCA) DBD by taking direct force measurements, velocity profiles, and time-resolved electrical and optical measurements [44]. Negative DCA electrode bias led to modest improvements before the onset of sliding discharge and a counter jet at the DCA electrode, canceling the gains from positive ion acceleration. These opposing wall jets collide, creating a wall-normal thrust [43]. In contrast, in some cases, a positive DCA bias monotonically increased thrust by more than twofold. An oscillating residual charge interaction mechanism is identified to explain the increase in horizontal thrust, in which the acceleration of positive ions is augmented by the attraction from the residual (negative) charge. While the results of the DCA-DBD are promising, their practical implementation requires AC and DC power supplies and is challenging for large-scale DBD arrays.

The E-field augmentation by an AC electrode can be more robust as it can utilize both E-field magnitude and phase shift control. Previous research described DBD arrays implementing several AC electrodes in series [21]. DBD arrays increase the total thrust; however, the subsequential exposed electrode can produce a counter wall jet (sometimes called a cross-talk phenomenon), limiting the overall system's efficiency [45]. In one of the earliest attempts to minimize DBD cross-talk, Benard *et al.* [46] proposed an array of three-electrode DBD actuators with two embedded electrodes. Later works found that alternating AC and ground electrodes reduced the cross-talk and continuously accelerated EHD flow. Debien *et al.* [22] reported EHD jet velocities up to ~ 10.5 m/s with a four-stage DBD array. The wire-to-planar DBD array in Debien *et al.* [22] demonstrated better mechanical performance than the planar-to-planar electrodes DBD array of Benard *et al.* [46]. Sato *et al.* [47] studied multi-stage DBD-driven flow at low AC voltages. Their DBD configuration with alternative HV-ground electrodes was similar to Debien *et al.* [22]; however, the system used a nanosecond pulsed (NP) DC waveform.

This study investigates AC-augmented (ACA) DBD actuators and outlines the optimization strategy for maximizing horizontal thrust. We utilize time-resolved electrical and optical measurements to optimize DBD performance as a function of electrical inputs, such as AC voltage, phase shift, and geometrical spacing. Time-resolved electrical and plasma discharge characteristics relate to time-averaged thrust and wall jet measurements, providing insights into the complex dynamics of ACA-DBD systems.

## 2. Experimental Setup and Diagnostics

### 2.1 AC-Augmented DBD Actuator

The schematic of the experimental setup for the AC-augmented (ACA) DBD actuator is shown in Figure 1. The three electrodes are identified as the DBD active electrode, the DBD embedded electrode, and the third or AC-augmented electrode. The discharge is generated with a high-voltage AC signal between the active and embedded electrodes. A 3.175 mm quartz plate separates the active and embedded electrodes. The active electrode (0.07 mm thick, 15 mm long, and 110 mm wide) is flush-mounting copper tape to the top of the dielectric surface. The embedded electrode (0.07 mm thick, 25 mm long, and 110 mm wide) is a copper tape flush mounted to the backside of the dielectric surface. The embedded electrode is encapsulated with a thick 1 mm Kapton layer (~7700 VPM @ 25°C) and silicone rubber tape (3.175 mm) to ensure no backside discharge. There is no overlap between the active and embedded electrodes. The ACA electrode is made of copper tape (0.07 mm thick, 15 mm long, and 110 mm wide) and is mounted downstream of the active electrode on the air-exposed side. For this study, the gap (L) was set at 10 mm and 25 mm. All tests were conducted in quiescent atmospheric air, T ~ 18-23 ºC, with relative humidity (RH) of ~30-50%.

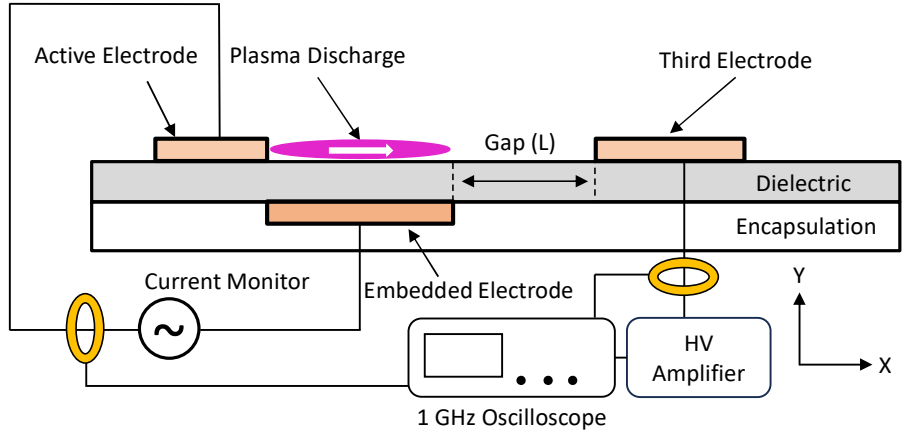

**Figure 1.** Experimental setup. The active and third electrodes are flush-mounted onto quartz dielectric. The embedded electrode is mounted on the back side of the dielectric layer and encapsulated with polyimide and silicone rubber layers. The active and embedded electrodes are connected to a custom power supply, while a Trek 40/15 HV amplifier powers the third electrode. The current is measured with a Pearson 2877 probe and a Tektronix DPO 7254C oscilloscope.

The active and embedded electrodes are connected to a custom high-voltage power supply similar to that of Thomas *et al.* [35]. Briefly, the custom power supply comprises a Siglent SDG1032X signal generator, a Crown XLi 3500 power amplifier, and two custom high-voltage transformers (Corona Magnetics) rated up to 35 kV peak-to-peak voltage ($V_{pp}$) each. The secondary winding of each transformer can be set out-of-phase from the other to reach $V_{pp}$ = 70 kV between the electrodes, or one of the transformers can be used with the embedded electrode connected to an electrical ground. In our test, we have not seen any difference in time-averaged results when a DBD is powered by a single transformer vs. two transformers with matching $V_{pp}$. The transformers connected to active and embedded electrodes were operated out-of-phase to obtain a greater voltage amplitude. This approach is analogous to a floating ground appropriate for a battery-powered DBD actuator. All tests in this study use a fixed 2 kHz sine wave with voltage across the active and embedded electrode between 25 to 40 kV peak-to-peak. The output voltage applied to the active electrode is monitored with a Tektronix P6015A high-voltage probe. The ACA electrode is connected to a Trek 40/15 high-voltage amplifier (0 to ± 40 kV peak voltage) and is excited with a 2 kHz AC sine wave with varying voltage and phase offset. The signal generator (Siglent SDG1032X) controls the input to the custom power supply and the Trek amplifier. The voltage to the ACA electrode is monitored directly from an output on the Trek amplifier. The voltage across the active and embedded electrodes will be referred to as $V_{DBD}$, while the augmenting voltage applied to the third electrode will be referred to as $V_{ACA}$. The third electrode phase shift ($\Phi$) is set relative to the active air-exposed DBD electrode, see Figure 2**.**

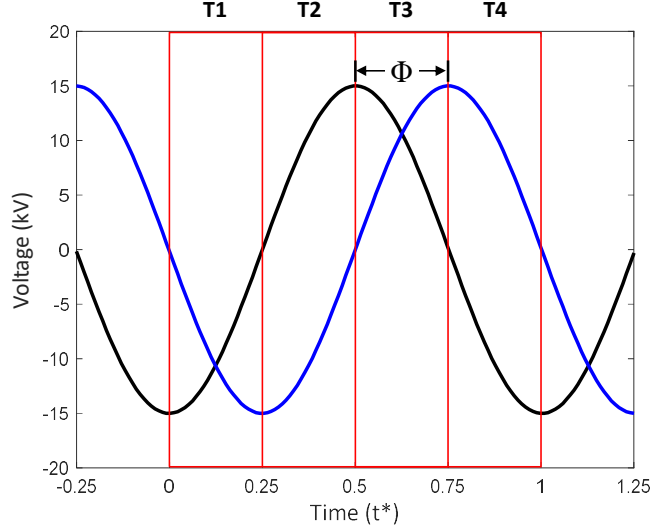

**Figure 2.** T1 - T4 quadrants of phase-resolved CCD imaging defined relative to $V_{DBD}$. The $V_{ACA}$ (blue) is phase-shifted by $\Phi$ from $V_{DBD}$ (black). The T1 and T2 quadrants capture discharges on the voltage-rising cycle associated with streamer discharge, while the T3 and T4 quadrants capture discharges on the voltage-falling cycle associated with glow discharge.

## 2.2   Electrical Discharge Measurements

DBD actuators are traditionally characterized by their electrical current or power usage. The electric current is a superposition of a capacitive and a discharge current. The discharge current is associated with the total number of ions generated during plasma microdischarges - a series of short current spikes [22]. The DBD active and third electrodes' current was measured using a 200 MHz bandwidth non-intrusive Pearson 2877 current monitor with a rise time of 2 ns resolution. The current was recorded using a Tektronix DPO 7254C oscilloscope (500 MHz bandwidth) with a sampling rate of 1 GS/s, essential for capturing individual discharges with a typical duration of ~30 ns [48]. The high bandwidth and the sampling rate minimize the noise during the current measurements and can be used to compute the time-averaged electrical power [52]. The capacitive current was identified to determine the currents associated with the plasma micro-discharges. Some previous reports have resolved the capacitive current through analytical methods [49, 50], while others have identified and removed the capacitive current through signal processing, including low-pass filters or Fast-Fourier Transform (FFT) [22, 48, 51]. In this work, the capacitive current was identified by the first 19 nodes of the Fourier series [18]. These nodes have a frequency up to ~ 5 times the voltage frequency and are close to that of the expected capacitive current.

Non-intrusive measurement of the electrical current allows computing the time-averaged electrical power consumption of the DBD actuator and the ACA electrode:

$$W_{elec} = f_{AC} \int_{t^*=0}^{t^*=1} V(t) * I(t)\, dt, \qquad (1)$$

where $f_{AC}$ is the frequency of the applied voltage in Hz, and $V(t)$ and $I(t)$ are the voltage and current at each point in the period. The normalized time ($t^*$) represents a single AC cycle. We compute the average, resulting power from at least ten separate voltage periods to reduce the noise impact. This approach to calculating power consumption differs from Lissajous curves that require the introduction of a capacitor between the embedded electrode and the power supply [21]. Since total power consumption is primarily driven by capacitance, other works, such as those by Kriegseis *et al.* [52], used Lissajous curve analysis to approximate the effective DBD capacitance with and without discharge present. The authors reported that both methods yield similar results; however, the overall current and discharge current were more consistent in the high bandwidth non-intrusive current monitors method than in the cases when the high bandwidth voltage probes were used for Lissajous curves.

## 2.3 Plasma Visualization

Time-integrated and phase-resolved optical images were taken to visualize the physical plasma discharge. Time-integrated photos were taken using a 24 MP Nikon D750 DSLR camera with a 200 mm macro lens (f/4.5). Each image is captured with a 20 ms exposure time, such as at $f_{AC}$ = 2kHz, which corresponds to 40 voltage cycles. This work uses a high-speed monochromatic CCD camera (Vision Research Phantom V12.1) coupled to a 200 – 550 nm UV intensifier lens (Specialised Imaging SIL3) to resolve the plasma propagation during each voltage cycle. In atmospheric air DBD discharge, all (visible and UV) light emissions are in the 300 - 450 nm range [53, 54]. For a 2 kHz frequency voltage cycle, the CCD and intensifier trigger is set to 8,000 Hz with a 120 μs CCD shutter and 105 μs intensifier shutter settings to divide the voltage cycle into four equal parts [22]. These quadrants, T1 through T4, represent the two halves of the voltage-rising cycle and two halves of the voltage-falling cycle (Figure 2). The CCD camera and intensifier are triggered by an additional function generator (Siglent SDG1032X) synchronized with the waveform generator of the custom power supply that controls the DBD actuator. For each test, the DBD is powered first, then the CCD camera and intensifier are triggered to capture images 3 seconds after the maximum voltage amplitude point of the cycle.

## 2.4 Wall Jet and Thrust Characterization

The mechanical performance of the DBD system is characterized by the horizontal direct thrust and velocity flow field created by the DBD system. Similar to other works [35], the direct thrust is measured by vertically holding the plasma actuator system onto an analytical balance. In this configuration, plasma-induced flow is directed away from the balance, and the balance measures the reactive downward thrust. An analytical balance (Ohaus SPX223, 0 – 220 g range with 1 μN resolution) was used. A Faraday cage was placed around the scale to prevent electromagnetic interference (EMI) or electrostatic forcing on the scale, and the DBD actuator plate was mounted to a non-conductive acrylic stand that rested ~ 50 mm above the balance. Thin conductive DBS (Drawn Brazed Strand) wire (0.1 mm OD) was used for all wire connections to the electrodes to ensure that wire tension does not influence the thrust measurements. The shielding against electrostatic or EMI was verified by placing the plasma actuator assembly ~ 10 mm above the stand but without contact; the balance did not register any "phantom" force when operating the DBD actuator. Thrust data from the balance was transferred to a computer through a shielded USB cable with galvanic isolation to prevent the scale from being grounded. Note that tests without the data isolation yielded nearly identical results. All thrust data points were averaged over three 10-second measurements.

To measure the time-averaged x-velocity profile, we employed a custom-made glass pitot tube with a 0.6 mm ID and 0.8 mm OD, similar to the previous report [18]. Compared to traditional stainless steel pitot tubes, the glass tube minimizes interaction with the discharge. The pitot tube is mounted on an optical table and is controlled in the x- and y-axis by linear stages. The pitot tube was connected to an Ashcroft CXLdp differential pressure transmitter (0 – 25 Pa with 0.25% accuracy). The pressure transmitter outputs a 4 – 20 mA current, linear in its pressure range, and it is connected in series with a 1.5 kΩ resistor. The pressure within the pitot tube equilibrates nearly instantly when the flow condition changes. The voltage across the resistor is recorded for at least 20 seconds with a Fluke Hydra Data Logger II. Time-averaged velocity ($v$) is calculated based on the time-averaged pressure ($P$) using Bernoulli's equation with a calibration correction factor ($C$) obtained for each custom pitot tube expressed as

$$\Delta P = C\rho v^2, \qquad (2)$$

where $\rho$ is the fluid density. Our experiments' velocity measurements typically have a standard deviation < 0.03 m s$^{-1}$ over the sampling period. X-velocity measurements are taken at varying x- and y - positions downstream and upstream on the active electrode edge. At each x-position, the velocity profile is obtained from the surface to 10 mm above the plate at increments of 0.25 mm or 0.5 mm (at a higher location). At the lowest height and accounting for the pitot tube diameter in a shear flow [55], the first velocity point is captured at y = 0.5 mm. We assume a linear profile between the y = 0 mm no-slip condition and the y = 0.5 mm measurements for plotting purposes.

The electric and resulting flow fields are assumed to be two-dimensional for straight-edged electrodes. Considering the 2D control volume (with a spanwise unit length), integration of the velocity along a vertical profile yields the total mass flow rate per meter spanwise, $Q$:

$$Q_{system} = \rho \int_{y=0}^{y=h} U(y)dy, \tag{3}$$

where $U(y)$ is the velocity measured along the vertical profile at a constant x-location. Similarly, the system's total momentum can be found by integrating the square of the velocity along a vertical profile:

$$M_{system} = \rho \int_{y=0}^{y=h} U^2(y)dy. \tag{4}$$

Without freestream flow nor external force, the momentum produced by the DBD represents a fluid-derived thrust measurement and is affected by the viscous effects within the control volume. As a result, velocity-derived thrust can slightly overpredict the DBD forcing compared to the directly measured thrust on the flat plate. For this study, the DBD electrodes were positioned to minimize unnecessary downstream length, and the control volume downstream vertical edge is at x = 15 mm, similar to previous works [18]. The upstream and top sides of the control volume are respectively considered at x = −20 mm and y = 20 mm, where velocity is typically zero. Using this control volume analysis method, we find that the control volume force estimation typically matches the directly measured force measurement within ~10%. Finally, the mechanical power of the system ($W_{mech}$) can be computed by

$$W_{mech} = \frac{1}{2}\rho L \int_{y=0}^{y=\infty} U^3(y)dy. \tag{5}$$

The electromechanical efficiency of the plasma actuator can be calculated based on the electrical power, Eq.(1), and mechanical power, Eq.(5), as

$$\eta = \frac{W_{mech}}{W_{elec}}. \tag{6}$$

## 3. Results: Plasma and Electrical Characteristics

### 3.1 Plasma Visualization

#### 3.1.1 Time-Integrated Plasma Visualization

The time-integrated plasma discharge visualization as a function of DBD voltage and phase shift in the ACA electrode Φ = 0 - 180° are presented with the smaller gap length, L = 10 mm, to understand the plasma regimes and interaction between the electrodes. Figure 3 shows the plasma images at varying third electrode phase shifts with a fixed $V_{DBD}$ = 30 kV and 40 kV and $V_{ACA}$ = 28 kV. The DBD active electrode is located on the bottom side of each image, and the third electrode is located on the top side. A steady glow discharge that increases with voltage can be seen from the DBD active electrode edge over the embedded electrode, with the possibility of the early formation of filamentary streamers.

For Φ = 0 - 60°, a streaming plasma discharge between the third and embedded electrodes is seen at the tested voltages. This reverse discharge from the ACA electrode shows strong filamentary streamers that extend to the area above the embedded electrode, in contrast to a more uniform glow discharge at the DBD active electrode. Interestingly, when the AC voltage is applied between the ACA and embedded electrode with the active electrode not energized, a typical DBD glow is seen at the ACA electrode edge without strong filamentary streamers (not shown). One possible explanation is the primary DBD produces a dominating space charge, leading to ACA electrode/space charge interaction rather than ACA / embedded electrode due to the air gap between the two. The smaller gap (L = 10mm) leads to the formation of persistent streamers, which are stronger at the higher DBD voltage condition Figure 3 (b), as more ions are generated.

Approaching $\Phi = 180°$, at higher $V_{DBD}$ potentials ($V_{DBD} > 30$ kV) or at higher $V_{ACA}$, the primary plasma discharge is extended (Figure 3 (b)), which is similar to the "sliding" discharge when strong negative DC bias is applied to a third electrode [44, 56]. In the sliding regime, the third electrode produces a counter-forcing discharge that results in a counter-flow and the deflection of the primary DBD jet. To the authors' knowledge, this is the first report of sliding DBD with a downstream AC bias. Mechanical behavior similar to other sliding DBD studies is observed in the thrust (Figure 9) and velocity measurements (Figure 12), which can be used to optimize the spacing of downstream electrodes before diminishing or interfering effects occur.

In the larger gap experiments (L = 25 mm), no sliding discharge was observed at the tested range of $V_{ACA}$, $V_{DBD}$, or $\Phi$. At $\Phi = 0°$, few reverse streamer discharges were observed at higher potentials, similar to Figure 3; however, the overall discharge was more homogenous. This is mainly due to a weaker electric field and reduced space charge density between the electrodes.

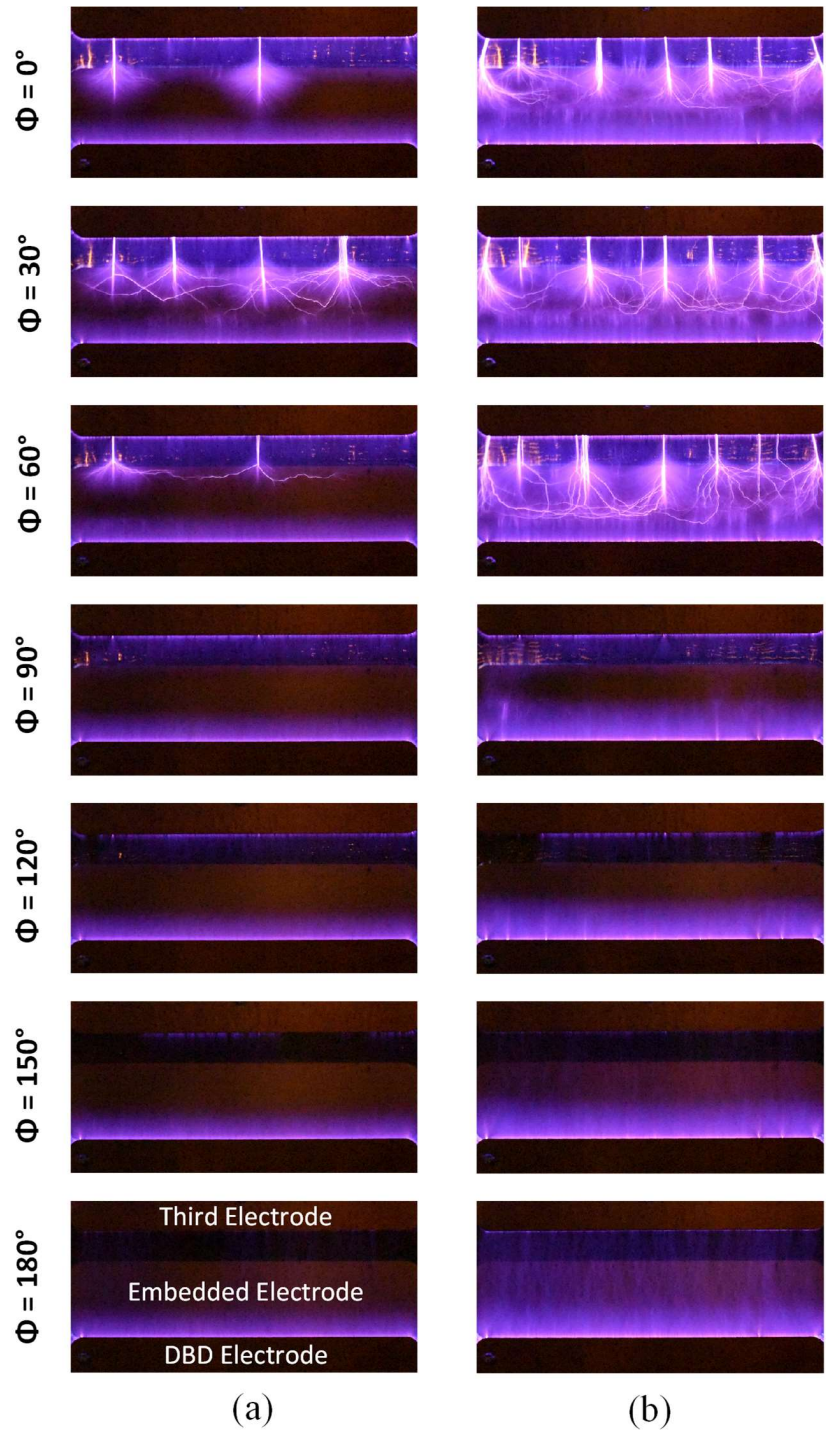

**Figure** 3. Time-integrated plasma emissions at varying ACA phase shift Φ, at $V_{DBD}$ = 30 kV (a) and 40 kV (b). The third electrode gap is 10 mm, and $V_{ACA}$ = 28 kV. The active DBD electrode is located at the bottom of each image; the primary EHD flow is directed upwards. The third electrode is located on the top of each image; the third electrode discharge is directed downwards.

### 3.1.2 Phase-Resolved Plasma Visualization

To complement time-integrated plasma visualizations, we obtained phase-resolved images to study the ACA effect on the plasma behavior. The baseline DBD case without ACA electrode at $V_{DBD}$ = 40 kV,

L = 10 mm is presented in Figure 4(a). The image orientation in Figure 4 is the same as in Figure 3, with the DBD electrode pair located at the bottom of each image. In the baseline case, the third electrode has a floating potential, and the actuator acts as a standard two-electrode DBD, allowing for a side-by-side comparison with ACA cases. With the passive third electrode, the plasma luminosity and length for all phases of the cycle increase with voltage magnitude. In voltage-rising quadrants (T1 and T2), the generation and growth of streamers are observed, while more homogenous glow discharge is seen during the voltage-falling quadrants (T3 and T4). The ends of each half cycle (T2 and T4) show higher luminosity and larger volume associated with the frequency and intensity of discharges, shown in Figure 5. Phase-resolved images of plate-to-plate DBD at 1 kHz in Debien *et al.* [22] agree well with the presented results and show that streamers in the positive-going half-cycle propagate further than the glow discharge of the negative-going half-cycle.

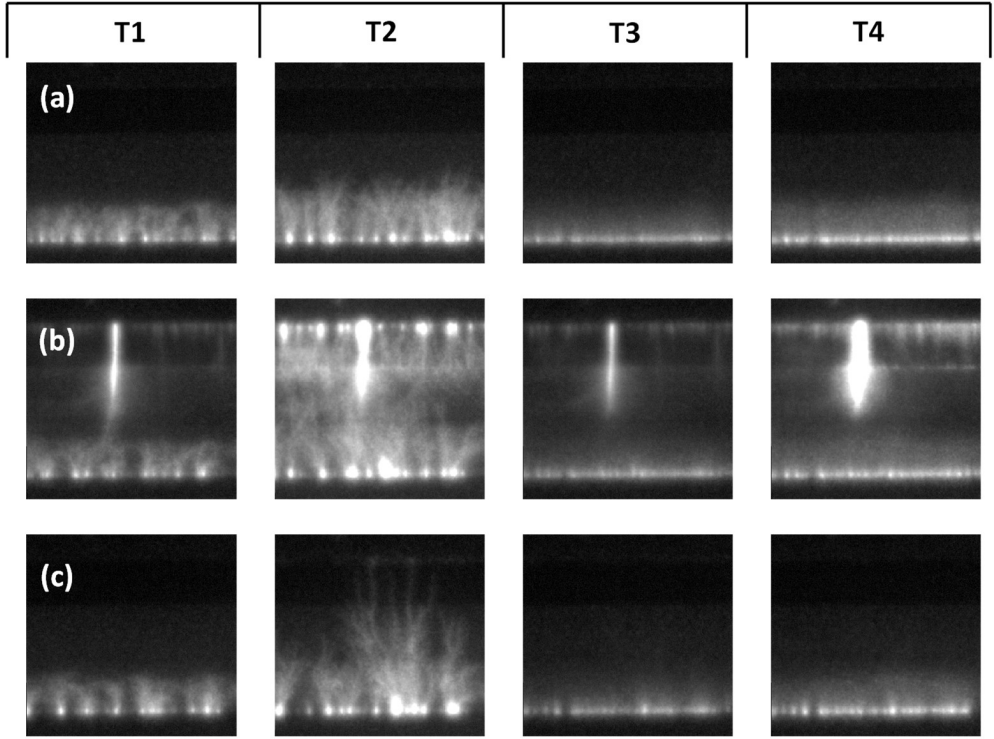

**Figure** 4. Phase-resolved plasma extension with a L = 10 mm gap at $V_{DBD}$ = 40 kV without a $V_{ACA}$ (a), with $V_{ACA}$ = 28 kV / Φ = 0° (b), and $V_{ACA}$ = 28 kV / Φ = 180° (c).

Adding ACA voltage leads to the strong interaction of the primary DBD plasma; these interactions depend on the electric field strength and phase shift. Figure 4 shows the phase-resolved images for L = 10 mm gap, Φ = 0° and Φ = 180°, $V_{DBD}$ = 40 kV with $V_{ACA}$ = 28 kV; these correspond to the time-averaged images in Figure 3. The reverse discharge on the ACA electrode for Φ = 0° (Figure 4 (b)) shows two opposing DBD discharges and the formation of persistent streamers present over several hundred cycles before moving along the electrode edge. Prior reports suggest that the residual surface charges can cause the onset of strong filamentary discharge [57]. Here, the discharge on the ACA electrode is similar to the primary electrode pair, e.g., in the T2 phase, several filamentary streamers extend to the embedded electrode, and a glow discharge is seen during the T3 and T4.

The out-of-phase ACA (Φ = 180°) case has the most noticeable difference during the T2 phase. Figure 4 (c) shows that for $V_{ACA}$ = 28 kV, the plasma extends to the third electrode, which is confirmed by negative discharge voltage on the ACA electrode, Figure 5. A significant reduction in the tangential momentum injection is observed during sliding discharge (Figure 9). Note that the sliding occurs during the T2 part of the cycle with a near out-of-phase shift between the exposed electrodes ($\Phi \sim 180°$). This condition corresponds to (i) a high primary discharge current, i.e., a high concentration of positive ions, and (ii) a high electric potential that drives the ions to the negatively charged ACA electrode. In the DBD-ACA system, the sliding discharge does not occur with negative species (T3 and T4) or when exposed electrodes are close to being in phase ($\Phi \sim 0°$).

## 3.2 Electrical Characteristics

Here, we present time-resolved electrical and discharge characterization as a function of DBD voltage and the ACA phase shift $\Phi = 0°$ and $180°$. Figure 5 shows a set of voltage and current graphs on the active DBD and ACA electrodes with an L = 10 mm gap. The electrical measurements for $V_{DBD} = 40$ kV and $V_{ACA} = 32$ kV complement the plasma emissions measurements, providing insight into the interaction between the charged species generated in the primary DBD plasma and E-filed modulated by the ACA electrode.

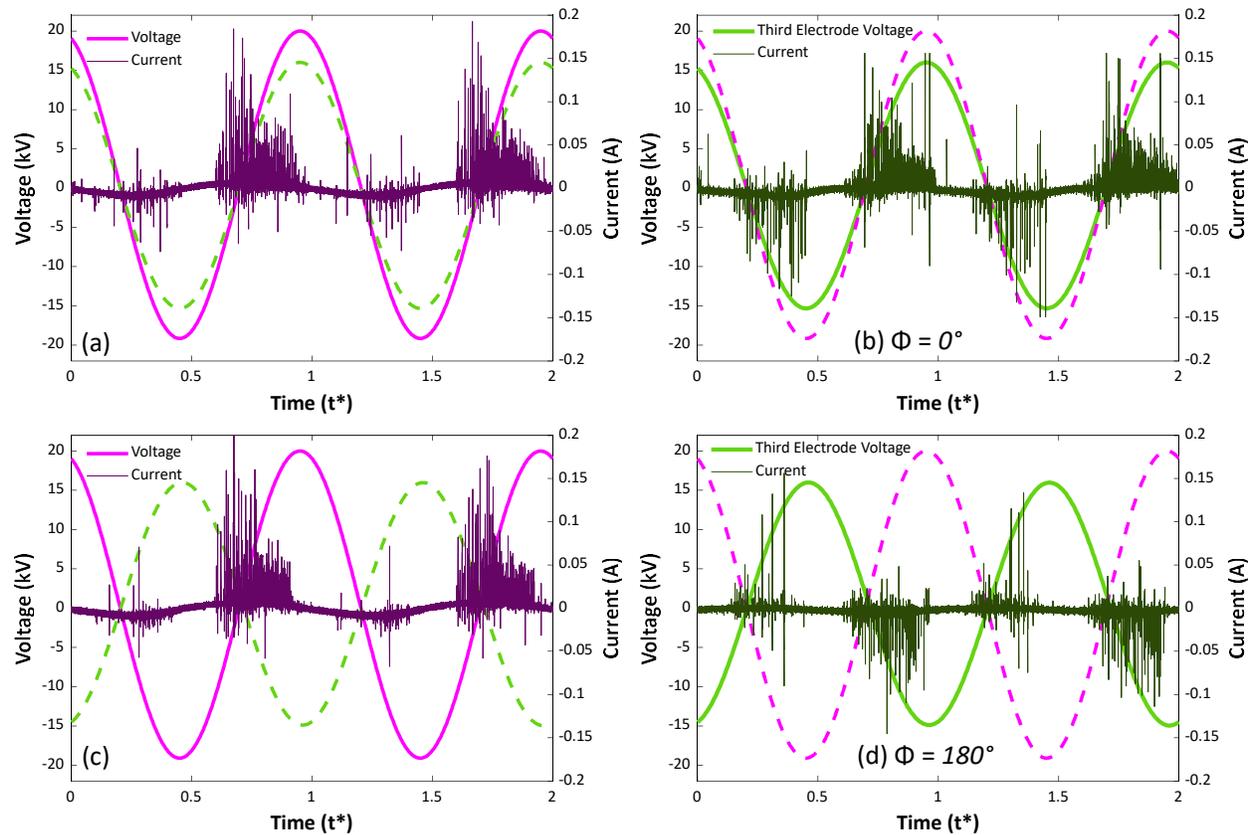

**Figure** 5. Instantaneous voltage and current of the primary DBD with gap distance L = 10 mm; smooth magenta line - primary DBD voltage, $V_{DBD} = 40$ kV; Smooth green line - ACA voltage, $V_{ACA} = 32$ kV. The thin purple line is the current on the primary, and the thin green line is the current through the ACA electrode. The dashed lines are provided for reference of the phase shift. (a, b) $\Phi = 0°$; (c, d) $\Phi = 180°$.

In the positive-going cycle, streamer discharges on the primary DBD, indicated by the positive discharge current spikes, can reach ~ 200 mA, Figure 5 (a, c). In the negative-going cycle, lower amplitude yet consistent, negative discharge current peaks are associated with the glow discharge related to the production of the negative species near the exposed electrode [21]. The compact spacing case, L =10 mm at $\Phi = 0°$ case, shows a significant positive discharge current on the third electrode at the same time as the primary DBD (Figure 5 (a, b)), suggesting that the opposing discharges on the third electrode lead to the space charge saturation and persistent streamer formation (Figure 4 (b)). In contrast, Figure 5 (c, d) shows the out-of-phase ($\Phi = 180°$) case with a negative discharge current on the ACA electrode, indicating the electrons' emissions from the third electrode toward the positive space charge. With sufficient E-field strength, such as high ACA voltage, the plasma extension leads to sliding discharge in the T2 phase (Figure 4 (c)). The negative discharge on a third electrode during sliding discharge was previously observed with a downstream high-voltage DC electrode [44, 56] and is consistent with our observations.

The effect of phase shift ($\Phi$) on the positive and negative discharge currents from the third electrode at L = 10 mm with $V_{DBD}$ = 25 kV and 40 kV are shown in Figure 6. The positive discharge current ($I_{dis}^+$) and negative discharge current ($I_{dis}^-$) are plotted as a function of $\Phi$. This maximum $I_{dis}^+$ indicates the reverse DBD (counter flow) associated with the saturated streamers, shown in Figure 3. In the stronger $V_{DBD}$ = 40 kV case (Figure 6,b), the ACA electrode experiences a reverse discharge at $\Phi = 0°$ with a maximum $I_{dis}^+$ = 10.5 mA/m and $I_{dis}^-$ = 3.0 mA/m. As $\Phi$ is increased, the reverse streamer discharge decreases. The onset of sliding discharge at $\Phi = 150°$ and $\Phi = 180°$ is reflected by an increase in $I_{dis}^-$. For low cases $V_{DBD}$ = 25 kV (Figure 6, a), the ACA electrode does not induce sliding discharge; thus, the ACA electrode does not trigger the reverse discharge even at $\Phi = 180°$. For these conditions, the ACA electrode does not contribute significant discharges to the system or modify the primary discharge; its primary contribution is augmenting the space charge produced by the primary DBD. In the stronger E-field scenario (L = 10 mm, $V_{DBD}$ = 40 kV), the third electrode produces significant discharge currents: positive in the opposing DBD discharge $\Phi = 0°$ (discharge power, $P_{DBD}$ = 193.9 W/m) and negative during the sliding discharge $\Phi = 180°$ ($P_{ACA}$ = 48.7 W/m). Note that the thrust values are lower in these conditions than during the optimal DBD-ACA operation, see Figure 9.

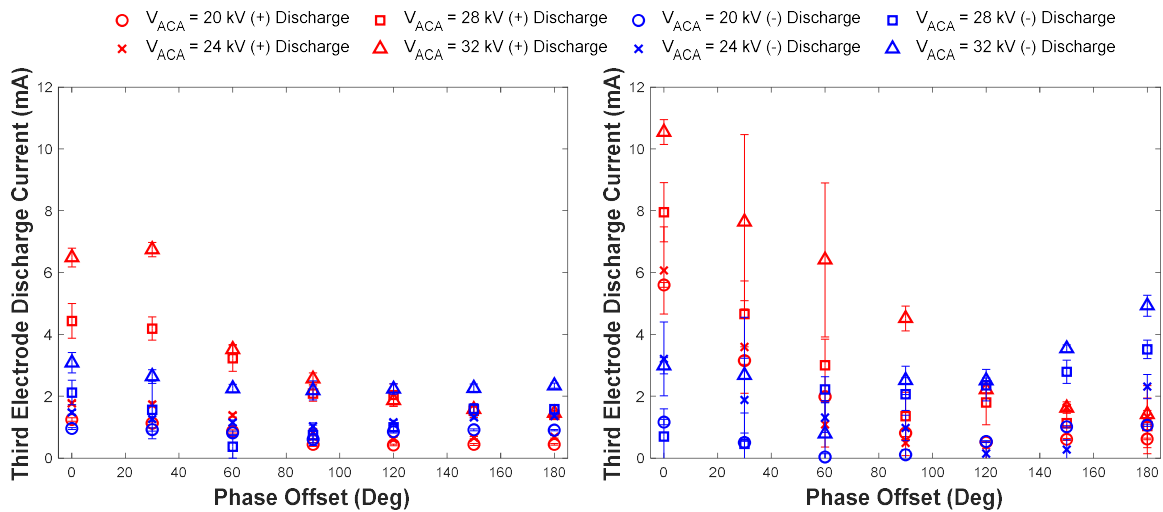

**Figure 6.** Third electrode positive and negative discharge current at varying phase shift and $V_{ACA}$ with L = 10 mm spacing (a) $V_{DBD}$ = 25 kV and (b) $V_{DBD}$ = 40 kV. The positive and negative discharge current is determined similarly to previous works [18]. The standard deviation is determined across at least 10 voltage cycles. Two standard deviation error bars are plotted.

The DBD-ACA actuator performance can be optimized by varying temporal E-field strength through primary and ACA voltage amplitude, phase shift, or electrode spacing. In the experiments at larger gap spacing, the electrical characteristics of the primary DBD were mainly independent of the potential and phase shift of the ACA electrode. At $V_{DBD}$ = 40 kV with a varying $V_{ACA}$ potential and $\Phi$, the total primary discharge current does not change significantly from its baseline of 11 mA/m. The electrical power usage of the DBD at $V_{DBD}$ = 40 kV across the range of $\Phi$ approximately remained constant at approximately 129 W/m.

Figure 7 shows the voltage and current graphs for DBD—ACA electrodes with 25 mm spacing, VDBD = 25 kV and 40 kV, similar to Figure 5. Applying the same VACA amplitudes weakens the E-field between the electrodes, resulting in significantly lower discharge currents compared to L = 10 mm. For Φ = 0°, the electrical characteristics of a reverse DBD are not observed, and for Φ = 180°, there is no visible sliding discharge between the exposed electrodes.

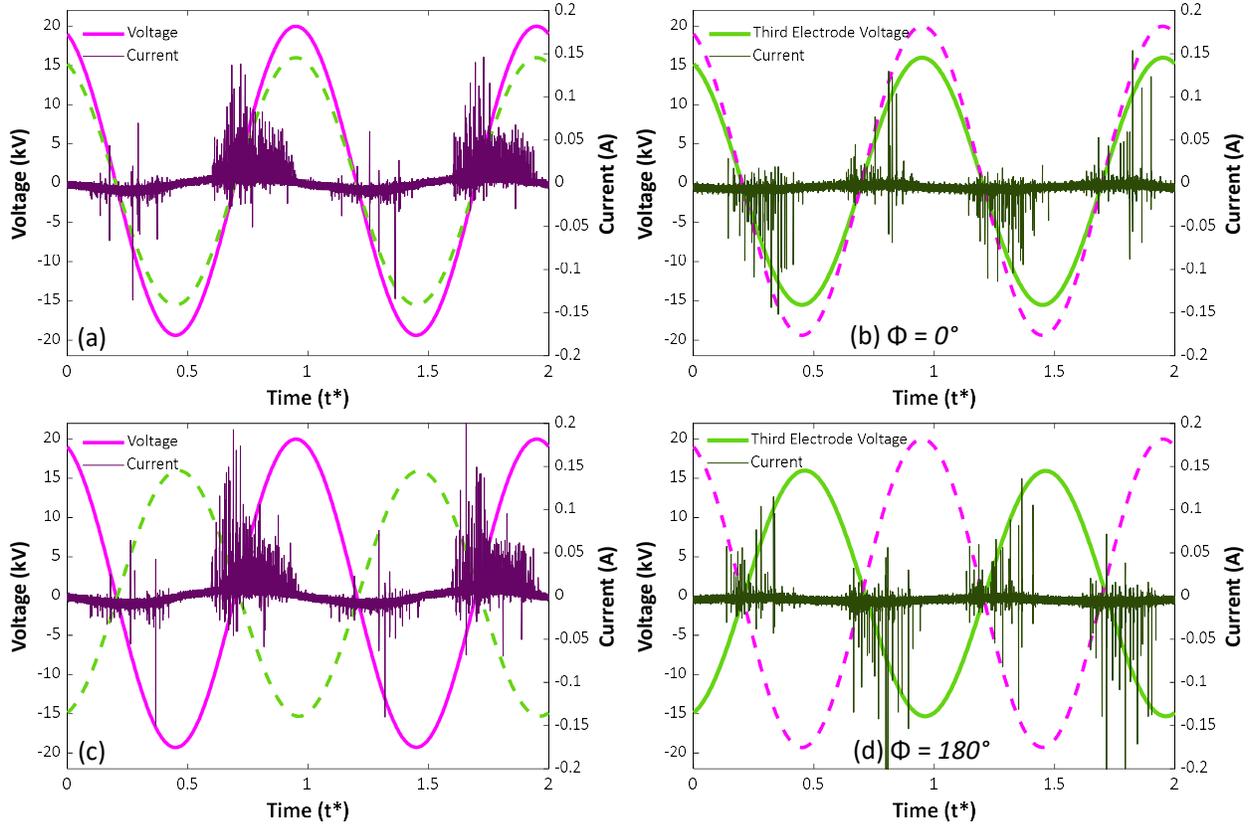

**Figure** 7. Instantaneous voltage and current of the primary DBD with gap distance L = 25 mm; magenta line = primary DBD voltage, $V_{DBD}$ = 40 kV; green line = ACA voltage, $V_{ACA}$ = 32 kV. The thin purple line is the current on the primary, and the thin green line is the current through the ACA electrode. The dashed line is provided for reference of the phase shift. (a, b) Φ = 0°; (c, d) Φ = 180°.

Integration of the discharge currents is shown in Figure 8. The effect of Φ on the positive and negative discharge current from the third electrode, $V_{ACA}$, at L = 25 mm with $V_{DBD}$ = 25 kV and 40 kV. With no reverse nor sliding discharge at L = 25 mm, significantly lower positive and negative discharge current is seen on the ACA third electrode. At the extended spacing, the third electrode discharge characteristics become primarily independent of the phase difference from the DBD; however, the discharge current is still dependent on $V_{ACA}$ potential. The third electrode thus mainly interacts with the unsteady plasma dynamics and the space charge, accelerating generated ions downstream from the DBD. The discharge current and total power usage for the primary DBD electrode for all Φ and $V_{ACA}$ conditions for L = 25 mm is similar to the baseline DBD for the L = 10 mm and L = 25 mm cases.

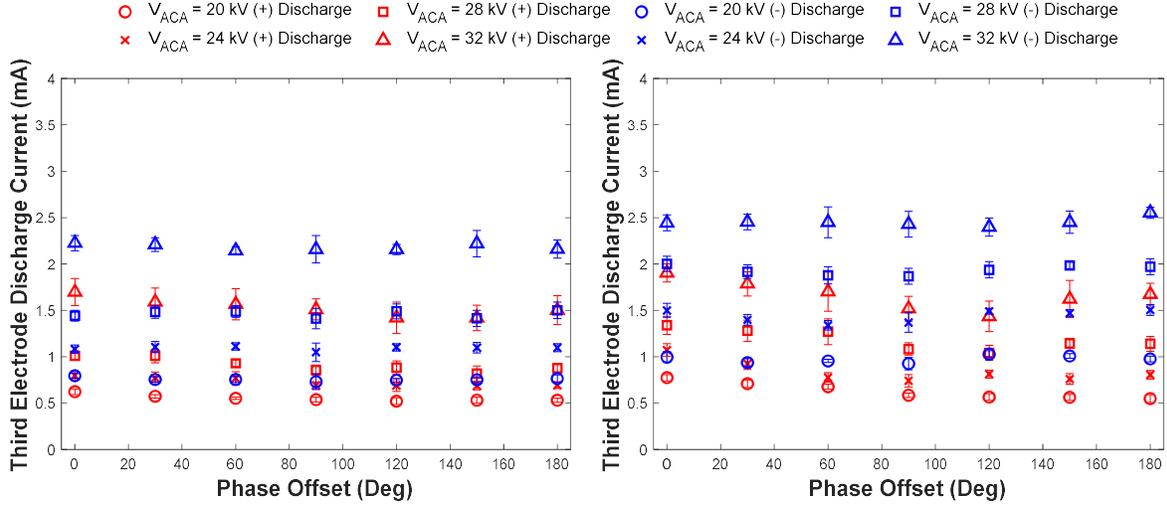

**Figure 8**. Third electrode positive and negative discharge current at varying phase shift and $V_{ACA}$ with L = 25 mm spacing at $V_{DBD}$ = 25 kV (a) and $V_{DBD}$ = 40 kV (b). The standard deviation is based on at least 10 voltage cycles. Two standard deviation error bars are plotted.

With the weaker E-field, the discharge power of the ACA electrode is significantly lower compared to the L = 10 mm case. In this configuration (L = 25 mm), the third electrode power use is at a maximum of 36.3 W/m for $\Phi = 60°$ and at a minimum of 21.9 W/m when $\Phi = 180°$ at $V_{ACA}$ = 32 kV. A summary of the power usage of the DBD and ACA electrodes is presented below in Table 1. The power usage is computed using Eq.(1) and is averaged across at least 10 voltage cycles.

**Table 1.** DBD and ACA electrode power usage at L = 10 mm and L = 25 mm.

| $V_{DBD}$ (kV) | $V_{ACA}$ (kV) | L (mm) | $\Phi$ (°) | $P_{DBD}$ (W/m) | $P_{ACA}$ (W/m) |
|---|---|---|---|---|---|
| 40 | 32 | 10 | 0 | 128.9 | 193.9 |
| 40 | 32 | 10 | 180 | 122.8 | 48.7 |
| 40 | 20 | 10 | 0 | 121.8 | 63.6 |
| 40 | 20 | 10 | 180 | 115.7 | 15.3 |
| 40 | 32 | 25 | 0 | 121.3 | 28.7 |
| 40 | 32 | 25 | 180 | 129.7 | 21.9 |
| 40 | 20 | 25 | 0 | 131.4 | 4.0 |
| 40 | 20 | 25 | 180 | 128.5 | 3.4 |

## 3.3 Mechanical Characteristics

### 3.3.1 Thrust

Figure 9 below shows the time-averaged thrust of the DBD-ACA actuator at L = 10 mm, f = 2 kHz, $V_{DBD}$ = 25 kV - 40 kV range, and $V_{ACA}$ = 20 kV - 32 kV with a varying phase shift. The typical standard deviation per point is <1 mN/m over the sampling period. The horizontal forcing is seen to generally increase from its baseline performance to a maximum at $\Phi = 180°$ and decrease to a minimum at $\Phi = 0°$. In all cases, the two air-exposed electrodes operated in phase with produced lower horizontal thrust decreased due to a counter-forcing reverse DBD from the third electrode (Figure 3). The most significant increase from 31.8 mN/m (baseline) to 41.8 mN/m at $\Phi = 180°$ (31%) is observed at $V_{DBD}$ = 35 kV, $V_{ACA}$ = 24 kV (Figure 9(c)). The highest total thrust of 54.5 mN/m was at $V_{DBD}$ = 40 kV, $V_{ACA}$ = 24 kV (Figure 9(d)) at $\Phi = 150°$. At lower $V_{DBD}$, such as $V_{DBD}$ = 30 kV, an out-of-phase case ($\Phi = 180°$), the thrust increased by ~ 38%.

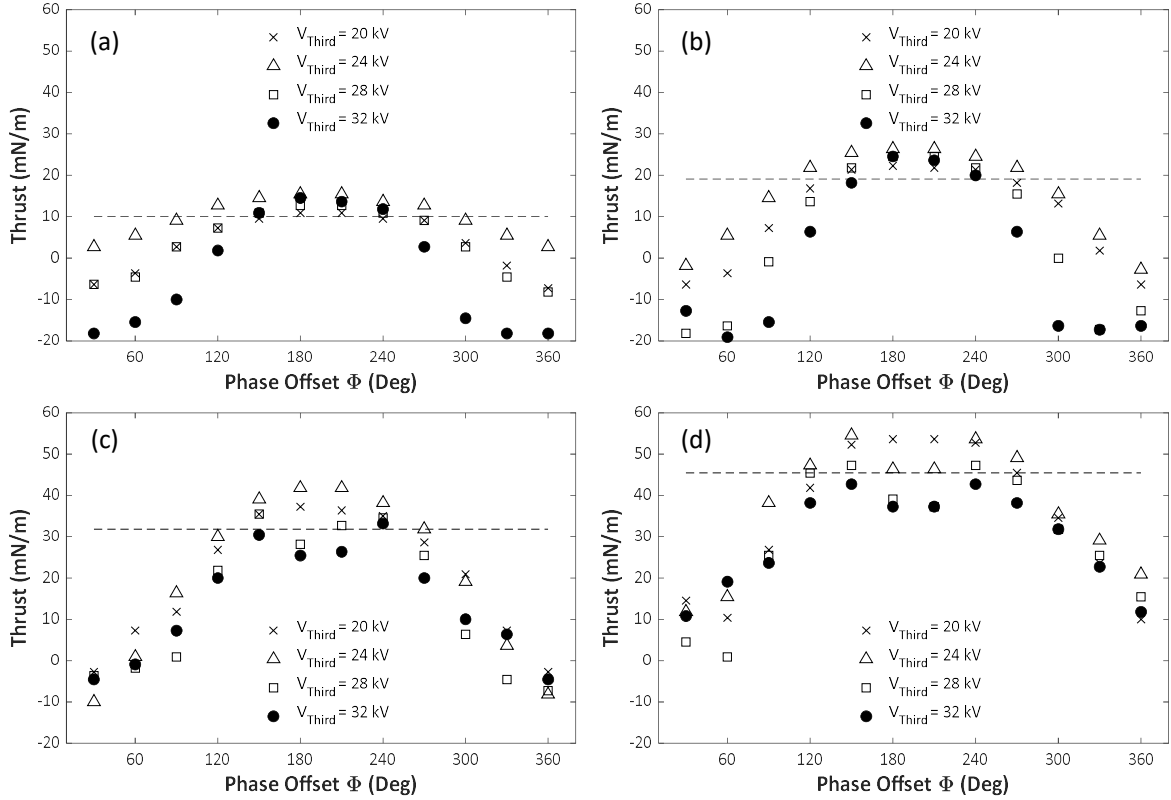

**Figure 9.** Horizontal Thrust at $V_{ACA}$ = 20 – 32 kV at 2 kHz with varying Φ with $V_{DBD}$ = 25 kV (a), 30 kV (b), 35 kV (c), and 40 kV (d) at 2 kHz. The distance from the embedded electrode to the third electrode is L = 10 mm. The DBD baseline thrust without a third electrode is a dashed line.

At the highest E-filed values, $V_{DBD}$ = 35 kV and 40 kV, $V_{ACA}$ > 24 kV, the thrust values dip below baseline at Φ = 150 - 240°. The decrease in horizontal forcing corresponds to sliding discharge cases exhibiting electrical characteristics similar to the DC-augmented sliding discharge [44]. In these cases, while the high electric field accelerates the positive ions generated at the DBD, negative electrical discharges from the third electrode to the space charge exert a backward force, diminishing the overall horizontal thrust. This observation is supported by discharge current measurements on the third electrode (Figure 6) and the time optical images of sliding discharge (Figure 3).

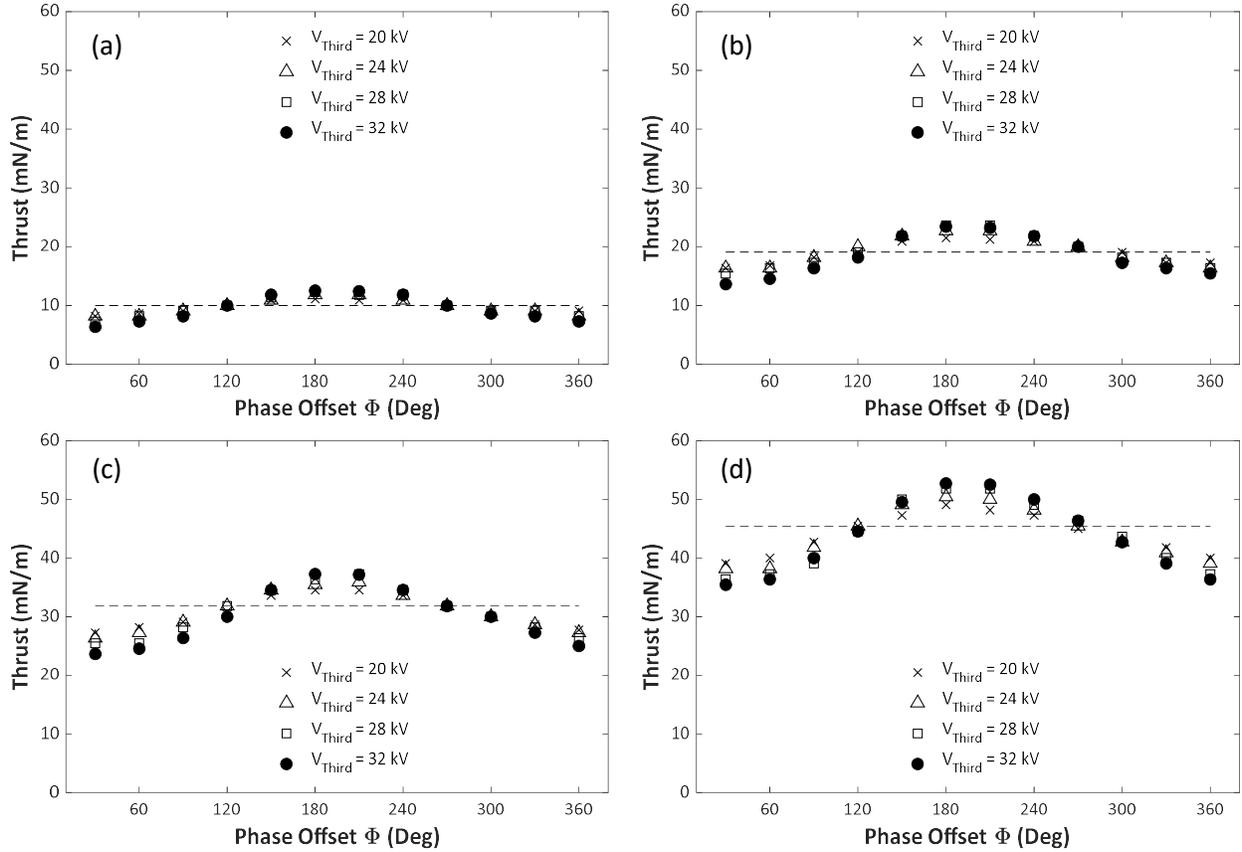

**Figure 10.** Thrust at $V_{ACA}$ = 20 – 32 kV at 2 kHz with varying phase shift. $V_{DBD}$ = 25 kV (a), 30 kV (b), 35 kV (c), and 40 kV (d). The third electrode gap L = 25 mm. The DBD baseline thrust without a third electrode is a dashed line.

Figure 10 shows the thrust with the third electrode gap L = 25 mm. In all cases, the maximum thrust is $\Phi$ = 180 - 210°, and the lowest thrust is at $\Phi$ = 0°. The maximum forcing with a single third electrode was seen when the DBD was operated at its maximum tested potential, $V_{DBD}$ = 40 kV, where the highest total thrust increased from 46 mN/m to 54 mN/m, an 18% increase. For $\Phi$ = 0°, the DBD thrust decreased from 46 mN/m to 36 mN/m, a 20% decrease. With a spacing of 25 mm between the embedded electrode and the third electrode, no sliding discharge was seen in all cases, supporting that sliding discharges decrease in thrust at the highest potentials in the L = 10 mm case (Figure 9). Compared to the L = 10 mm cases, the L = 25 mm condition has a slightly smaller thrust enhancement for the same $V_{DBD}$ due to the weaker electric field; however, the weaker electric field allowed for a higher $V_{DBD}$ before the onset streamers or sliding discharge. When $\Phi$ = 180° with $V_{DBD}$ = 40 kV with $V_{ACA}$ = 32 kV, the L = 25 mm spacing has a thrust of 53.6 mN/m compared to the 37.3 mN/m at L = 10 mm. A summary of the thrust measurements is presented in Table 2 for the $V_{DBD}$ = 40 kV cases.

**Table 2.** Total thrust measurement for $V_{DBD}$ = 40 kV at L = 10 mm and L = 25 mm cases.

| $V_{DBD}$ (kV) | $V_{ACA}$ (kV) | L (mm) | Φ (°) | Thrust (mN/m) |
|---|---|---|---|---|
| 40 | – | - | – | 45.5 |
| 40 | 24 | 10 | 150 | 54.5 |
| 40 | 28 | 10 | 180 | 39.1 |
| 40 | 32 | 10 | 180 | 37.3 |
| 40 | 24 | 25 | 180 | 50.5 |
| 40 | 28 | 25 | 180 | 51.8 |
| 40 | 32 | 25 | 180 | 53.6 |

*On the mechanism of plasma flow interaction in an AC-augmented system*: As demonstrated by optical and electric measurements, primary DBD discharge characteristics do not significantly change between the baseline and the ACA cases; thus, change in horizontal thrust is the result of the temporal manipulation of the species generated in the primary DBD by the ACA modulated E-field. Mechanistically, while the ACA may enhance the push-push DBD plasma/ flow interaction in the primary DBD, *performance improvements are primarily enabled by the third electrode's charge pull action*. The out-of-phase third electrode enhances the positive ion momentum transfer in the DBD positive-going cycle and the electron momentum or charge transfer in the DBD negative-going cycle. Prior work on temporal surface charge measurements comments that a negative surface potential in the negative-going cycle and positive surface potential in the positive-going cycle is dominated by physically charged ions on the dielectric surface [58]. Our recent work on the DCA-DBD also suggests that (i) the complex interaction of negatively charged species with neutral molecules and (ii) the oscillating behavior of residual surface and space are likely to play a role in the AC-augmented plasmas [44]. This work did not directly evaluate the relative contribution of the positive or negative species to thrust.

### 3.3.2 Wall Jet Characteristics

The EHD jet characterization confirms the thrust measurements and sheds insight into momentum generation in the DBD-ACA system. The conditions with the most significant thrust measurement change are the out-of-phase and in-phase ACA shifts. Thus, we compared the X-velocity profiles for these two cases and the baseline. For L = 10 mm with the third electrode 35 mm downstream of the active electrode, three positional measurements were taken at X = 15 mm, 25 mm, and 40 mm. These positions allow velocity characterization immediately after the DBD plasma region, X ~ 15 mm to 20 mm, and upstream and downstream of the third electrode. For L = 25 mm, the third electrode is 50 mm downstream of the active electrode; thus, the velocity profiles are taken at X = 15 mm, 25 mm, 40 mm, and 55 mm. The X-velocity profiles were also taken at Y = 1.0 mm, 2.0 mm, and 3.0 mm. Two standard deviation error bars are shown with each data point.

Figure 11 shows the baseline DBD and the DBD-ACA with in-phase and out-of-phase spacing at an L = 25 mm. Figure 11(a) and Figure 11(b) show the baseline (2-electrode DBD) with a typical DBD wall jet: $V_{max}$ = 5.68 m/s at X = 15 mm at the near-wall height of Y = 0.5 mm. Higher velocities can occur closer to or inside the plasma region; however, this region was not probed to prevent dielectric charging of the pitot tube. Viscous effects and surface/space charge interaction lead to momentum displacement downstream; this observation agrees with several previously published results in a quiescent environment, e.g., ref [21]. The directly measured thrust (45.5 mN/m) closely agrees with the velocity-derived thrust of 48.6 mN/m using equation (4). The momentum at the furthest measured distance of 55 mm shows < 10% loss due to viscous dissipation [8, 18]. Figure 11(a) and Figure 11(b) show the velocities for cases with activated ACA electrodes; the DBD profiles show slightly diminished EHD forcing Φ = 0° or enhanced Φ = 180°. Using the control, the velocity-derived thrust of the DBD with an in-phase third electrode is 39.8 mN/m, which agrees directly with the measured thrust of 36.4 mN/m (9%). Out-of-phase ACA velocity derived-thrust of 52.3 mN/m agrees with the directly measured thrust of 53.1 mN/m.

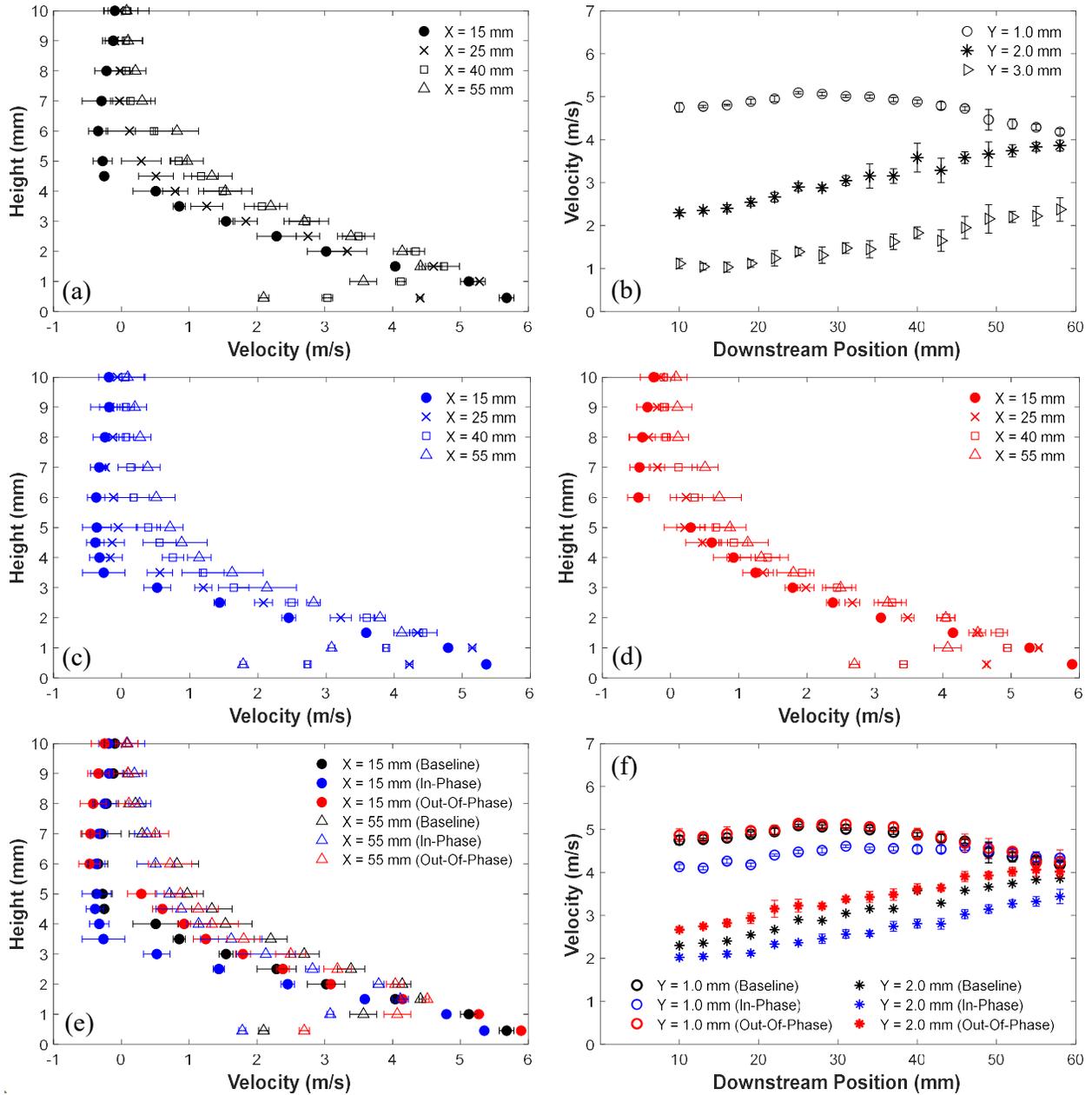

**Figure 11.** X-velocity profiles: $V_{DBD}$ = 40 kV, L = 25 mm; $V_{ACA}$ = 0 kV, (a) and (b). $V_{ACA}$ = 32 kV $\Phi$ = 0° (c) and $\Phi$ = 180° (d). The profiles are compared at X = 15 mm and X = 55 mm (e), y = 1 mm and y = 2 mm (f).

Figure 11(e) and Figure 11(f) compare the baseline profiles with the activated ACA. The out-of-phase cases show a stronger jet near the wall. At Y = 2.0 mm, the DBD jet, in all cases, shows nearly identical velocities. Notably, the three conditions show no sign of any additional or reverse forcing at the third electrode since the three conditions show similar profile changes, especially immediately before and after the third electrode. Thus, the modified DBD forcing regions appear near the primary DBD electrode. For stronger E-filed L = 10 mm case, the effect of ACA shows the diminishing impact on the velocity profiles for $\Phi$ = 0° (opposing jet) and $\Phi$ = 180° (sliding discharge); see Figure 12.

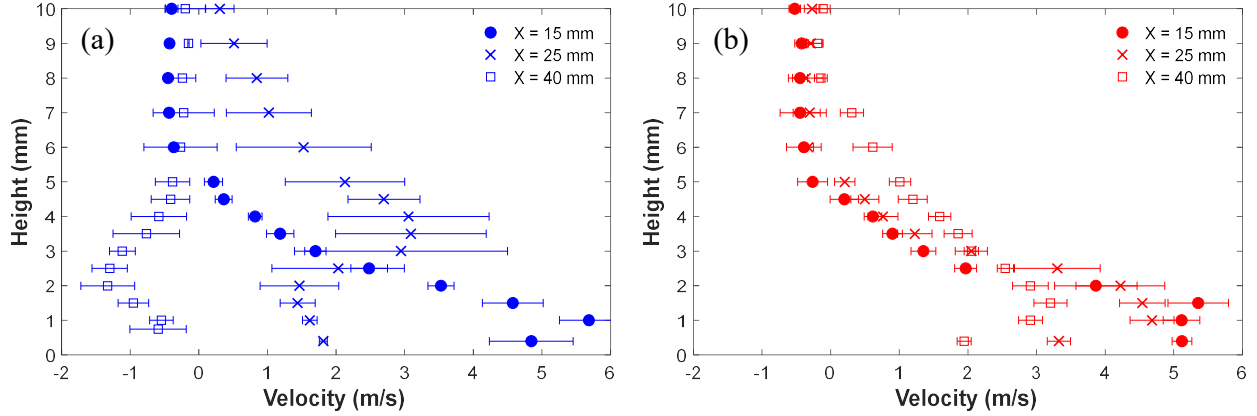

**Figure 12.** X-velocity profiles: $V_{DBD} = 40$ kV in-phase with $V_{ACA} = 20$ kV (a) and $V_{DBD} = 40$ kV out-of-phase with $V_{ACA} = 28$ kV (b) with a L = 10 mm third electrode gap.

When measuring profiles with $\Phi = 0°$ and a L = 10 mm gap, a $V_{ACA} = 20$ kV potential was used because higher potentials caused significantly more unsteadiness and nonuniformity, highlighting that the velocity field likely is no longer two-dimensional due to the strong reverse streamers, as shown in Figure **3**. The directly measured thrusts of the two conditions are shown in Figure 9. Figure 12(a) shows the collision of two jets with the reverse flow at X = 40 mm; however, the profile at X = 15 is similar to the baseline case. Momentum analysis of the $\Phi = 180°$ case (Figure 12(b)) shows lower forcing due to sliding discharge likely originating at the third electrode. In this case, the integrated momentum at x = 15 mm and 25 mm is 58.1 mN/m and 53.4 mN/m, respectively. However, the momentum at X = 40 mm is 31.2 mN/m. Since X = 40 mm is located directly after the edge of the third electrode, the drop in momentum is, therefore, from reverse forcing due to ACA discharge in the sliding DBD. The slight reverse forcing due to sliding discharge is supported by the measured negative discharge shown in Figure 5**,** corresponding to the phase-resolved positive sliding discharge shown in Figure 4(c). Due to the higher momentum at the X = 15 mm position, the authors believe the out-of-phase ACA still accelerates the positive ions from the primary DBD, as in the L = 25 mm case; however, this gain is lost by the reverse forcing from the sliding discharge. The L = 10 mm case highlights a possible spacing limit in multi-electrode systems for horizontal thrust as reverse discharge or sliding discharges may cause reverse forcing, leading to less efficient operation.

The mechanical power is calculated for $V_{DBD} = 40$ kV / $V_{ACA} = 32$ kV at $\Phi = 180°$ and L = 25 mm, using equation (6). The DBD actuator's power usage increased from the baseline of 0.133 W/m to 0.147 W/m with the out-of-phase third electrode. The efficiency of ACA-DBD is almost identical to that of the two-electrode DBD. The overall efficiency at the highest power condition is ~0.1 %, twice as high as reported in the literature (~ 0.05%) for dielectric of similar thickness [22]. Since the increase in momentum injection plateaus near $V_{ACA} = 32$ kV, weaker electrical cases may have greater efficiency. A summary of the electromechanical characteristics is shown in Table 3.

**Table 3.** Velocity-derived momentum, power, and efficiency of the three-electrode DBD system compared to the baseline two-electrode DBD. The momentum can be compared to the directly measured thrust.

| $V_{DBD}$ (kV) | $V_{ACA}$ (kV) | L (mm) | $\Phi$ (°) | Thrust (mN/m) | Momentum (mN/m) | $P_{mech}$ (W/m) | $P_{elec, total}$ (W/m) | $\eta$ (%) |
|---|---|---|---|---|---|---|---|---|
| 40 | – | – | – | 45.5 | 48.6 | 0.133 | 131.7 | 0.100 |
| 40 | 28 | 10 | 180 | 39.1 | 43.7 | 0.104 | 162.8 | 0.064 |
| 40 | 32 | 25 | 180 | 53.6 | 52.3 | 0.147 | 151.6 | 0.097 |

# 4. Conclusions

This is the first report of electromechanical characteristics of an AC-augmented DBD actuator. The three-electrode DBD actuator is characterized using time-integrated and time-resolved plasma visualizations, time-resolved current analysis, thrust measurements, and velocity profiles. The experimental results and the analysis provide insights into the DBD-ACA operation and can be used to optimize high-power density non-thermal plasma devices.

- The performance of an ACA-DBD actuator depends on the phase shift in voltage waveform between the primary DBD and the third electrode, the voltage amplitude, and the spacing between the DBD and the third electrode.
- Thrust, velocity profiles, and electrical current measurements show that an out-of-phase ACA enhances the primary DBD body forcing. The highest measured thrust with the out-of-phase ACA is 53.6 mN/m, $V_{max} \sim 6$ m/s, and electromechanical efficiency is $\sim 0.10\%$.
- The in-phase ACA ($\Phi = 0°$) can cause the formation of reverse discharge and flow. The out-of-phase ACA ($\Phi = 180°$) accelerates both the positive and negative species generated by primary DBD, enhancing momentum transfer to the neutral molecules.
- For $\Phi = 180°$ and with a sufficiently strong E-field, a sliding discharge occurs during the positive-going voltage cycle by inducing a negative discharge from the third electrode to the space charge. This phenomenon was previously only seen in DC-augmented DBD systems. The onset of sliding discharge is likely to be a critical spacing limiting factor in DBD arrays, limiting horizontal thrust and efficiency.

Mechanistically, while the ACA may enhance the push-push DBD plasma/ flow interaction mechanism of the primary DBD, the improvements in thrust are primarily due to the additional charge pull by the third electrode. While this work does not directly evaluate the relative contribution of the positive or negative species, future research should probe (i) the complex interaction of charged species with neutral molecules and (ii) the oscillating behavior of residual surface and space in the AC-augmented plasmas. These investigations should include time-resolved velocity and surface charge measurements. The findings of this work can be used as validation data for numerical model developments and optimization of DBD arrays.

# 5. Acknowledgments

The authors would like to thank Professor Alberto Aliseda for allowing the use of the high-speed Vision Research Phantom V12.1 CCD camera and Professor Uri Shumlak for allowing the use of the Specialised Imaging SIL3 UV intensifier.

# 6. References

[1] I. Adamovich *et al.*, "The 2017 Plasma Roadmap: Low temperature plasma science and technology," *Journal of Physics D: Applied Physics,* vol. 50, no. 32, p. 323001, 2017.
[2] E. Moreau, "Airflow control by non-thermal plasma actuators," *Journal of Physics D: Applied Physics,* vol. 40, no. 3, pp. 605-636, 2007, doi: 10.1088/0022-3727/40/3/S01.
[3] G. Fridman *et al.*, "Comparison of direct and indirect effects of non-thermal atmospheric-pressure plasma on bacteria," *Plasma Processes and Polymers,* vol. 4, no. 4, pp. 370-375, 2007.
[4] A. Tang, A. Ong, N. Beck, J. S. Meschke, and I. V. Novosselov, "Surface Virus Inactivation by Non-Thermal Plasma Flow Reactor," 2023.
[5] R. S. Vaddi, Y. Guan, and I. Novosselov, "Behavior of ultrafine particles in electrohydrodynamic flow induced by corona discharge," *Journal of Aerosol Science,* p. 105587, 2020.
[6] O. Ekin and K. Adamiak, "Electric field and EHD flow in longitudinal wire-to-plate DC and DBD electrostatic precipitators: A numerical study," *Journal of Electrostatics,* vol. 124, p. 103826, 2023.
[7] A. Michels, S. Tombrink, W. Vautz, M. Miclea, and J. Franzke, "Spectroscopic characterization of a microplasma used as ionization source for ion mobility spectrometry," *Spectrochimica Acta Part B: Atomic Spectroscopy,* vol. 62, no. 11, pp. 1208-1215, 2007.
[8] R. S. Vaddi, Y. Guan, A. Mamishev, and I. Novosselov, "Analytical model for electrohydrodynamic thrust," *Proceedings of the Royal Society A,* vol. 476, no. 2241, p. 20200220, 2020.
[9] H. K. Hari Prasad, R. S. Vaddi, Y. M. Chukewad, E. Dedic, I. Novosselov, and S. B. Fuller, "A laser-microfabricated electrohydrodynamic thruster for centimeter-scale aerial robots," *PloS one,* vol. 15, no. 4, p. e0231362, 2020.
[10] H. Xu *et al.*, "Flight of an aeroplane with solid-state propulsion," *Nature,* vol. 563, no. 7732, pp. 532-535, 2018.
[11] N. Gomez-Vega, H. Xu, J. M. Abel, and S. R. Barrett, "Performance of decoupled electroaerodynamic thrusters," *Applied Physics Letters,* vol. 118, no. 7, 2021.
[12] T. C. Corke, C. L. Enloe, and S. P. Wilkinson, "Dielectric Barrier Discharge Plasma Actuators for Flow Control *," *Annu. Rev. Fluid Mech.,* vol. 42, no. 1, pp. 505-529, 2010, doi: 10.1146/annurev-fluid-121108-145550.
[13] T. C. Corke, M. L. Post, and D. M. Orlov, "SDBD plasma enhanced aerodynamics: concepts, optimization and applications," *Progress in Aerospace Sciences,* vol. 43, no. 7, pp. 193-217, 2007, doi: 10.1016/j.paerosci.2007.06.001.
[14] Y. Guan, R. S. Vaddi, A. Aliseda, and I. Novosselov, "Analytical model of electrohydrodynamic flow in corona discharge," *Physics of Plasmas,* vol. 25, no. 8, p. 083507, 2018.
[15] Y. Guan, R. S. Vaddi, A. Aliseda, and I. Novosselov, "Experimental and numerical investigation of electrohydrodynamic flow in a point-to-ring corona discharge," *Physical Review Fluids,* vol. 3, no. 4, p. 043701, 2018.
[16] Z. Su, H. Zong, H. Liang, J. Li, and X. Chen, "Characteristics of a dielectric barrier discharge plasma actuator driven by pulsed-DC high voltage," *Journal of Physics D: Applied Physics,* vol. 55, no. 7, p. 075203, 2021.
[17] G. Touchard, "Plasma actuators for aeronautics applications-State of art review," *International Journal of Plasma Environmental Science and Technology,* vol. 2, no. 1, pp. 1-25, 2008.
[18] A. Tang, R. S. Vaddi, A. Mamishev, and I. V. Novosselov, "Empirical relations for discharge current and momentum injection in dielectric barrier discharge plasma actuators," *Journal of Physics D: Applied Physics,* vol. 54, no. 24, p. 245204, 2021.
[19] N. Benard, P. Noté, M. Caron, and E. Moreau, "Highly time-resolved investigation of the electric wind caused by surface DBD at various ac frequencies," *Journal of electrostatics,* vol. 88, pp. 41-48, 2017, doi: 10.1016/j.elstat.2017.01.018.

[20] E. Moreau *et al.*, "Surface dielectric barrier discharge plasma actuators," *ERCOFTAC Bulletin,* vol. 94, pp. 5-10, 2013.

[21] N. Benard and E. Moreau, "Electrical and mechanical characteristics of surface AC dielectric barrier discharge plasma actuators applied to airflow control," *Experiments in Fluids,* vol. 55, no. 11, p. 1846, 2014.

[22] A. Debien, N. Benard, and E. Moreau, "Streamer inhibition for improving force and electric wind produced by DBD actuators," *Journal of Physics D: Applied Physics,* vol. 45, no. 21, p. 215201, 2012.

[23] A. Hoskinson, L. Oksuz, and N. Hershkowitz, "Microdischarge propagation and expansion in a surface dielectric barrier discharge," *Applied Physics Letters,* vol. 93, no. 22, p. 221501, 2008.

[24] N. Benard, A. Debien, and E. Moreau, "Time-dependent volume force produced by a non-thermal plasma actuator from experimental velocity field," *Journal of Physics D: Applied Physics,* vol. 46, no. 24, p. 245201, 2013.

[25] M. T. Hehner, D. Gatti, and J. Kriegseis, "Stokes-layer formation under absence of moving parts—A novel oscillatory plasma actuator design for turbulent drag reduction," *Physics of Fluids,* vol. 31, no. 5, p. 051701, 2019.

[26] A. Duong, T. Corke, F. Thomas, and K. Yates, "Turbulent boundary layer drag reduction using pulsed-dc plasma actuation," *Journal of Fluid Mechanics,* 2019.

[27] M. Han, J. Li, Z. Niu, H. Liang, G. Zhao, and W. Hua, "Aerodynamic performance enhancement of a flying wing using nanosecond pulsed DBD plasma actuator," *Chinese Journal of Aeronautics,* vol. 28, no. 2, pp. 377-384, 2015, doi: 10.1016/j.cja.2015.02.006.

[28] N. Benard, J. Jolibois, and E. Moreau, "Lift and drag performances of an axisymmetric airfoil controlled by plasma actuator," *Journal of Electrostatics,* vol. 67, no. 2 3, p. 133, 2009, doi: 10.1016/j.elstat.2009.01.008.

[29] S. G. Pouryoussefi, M. Mirzaei, F. Alinejad, and S. M. Pouryoussefi, "Experimental investigation of separation bubble control on an iced airfoil using plasma actuator," *Applied Thermal Engineering,* vol. 100, pp. 1334-1341, 2016.

[30] L. Francioso, C. De Pascali, E. Pescini, M. De Giorgi, and P. Siciliano, "Modeling, fabrication and plasma actuator coupling of flexible pressure sensors for flow separation detection and control in aeronautical applications," *Journal of Physics D: Applied Physics,* vol. 49, no. 23, p. 235201, 2016.

[31] D. S. Drew and K. S. Pister, "First takeoff of a flying microrobot with no moving parts," in *Manipulation, Automation and Robotics at Small Scales (MARSS), 2017 International Conference on*, 2017: IEEE, pp. 1-5.

[32] R. S. Vaddi, Y. Guan, A. Mamishev, and I. Novosselov, "Analytical model for electrohydrodynamic thrust," *Proceedings A: Mathematical, Physical and Engineering Sciences,* vol. 476, no. 2241, 2020.

[33] D. Keisar, D. Hasin, and D. Greenblatt, "Plasma actuator application on a full-scale aircraft tail," *AIAA Journal,* vol. 57, no. 2, pp. 616-627, 2019.

[34] H. Akbıyık, H. Yavuz, and Y. E. Akansu, "Comparison of the linear and spanwise-segmented DBD plasma actuators on flow control around a NACA0015 airfoil," *IEEE Transactions on Plasma Science,* vol. 45, no. 11, pp. 2913-2921, 2017.

[35] F. O. Thomas, T. C. Corke, M. Iqbal, A. Kozlov, and D. Schatzman, "Optimization of Dielectric Barrier Discharge Plasma Actuators for Active Aerodynamic Flow Control," *AIAA Journal,* vol. 47, no. 9, pp. 2169-2178, 2009, doi: 10.2514/1.41588.

[36] J.-S. Yoon and J.-H. Han, "One-equation modeling and validation of dielectric barrier discharge plasma actuator thrust," *Journal of Physics D: Applied Physics,* vol. 47, no. 40, p. 405202, 2014.

[37] A. R. Hoskinson and N. Hershkowitz, "Differences between dielectric barrier discharge plasma actuators with cylindrical and rectangular exposed electrodes," *Journal of Physics D: Applied Physics,* vol. 43, no. 6, p. 065205, 2010, doi: 10.1088/0022-3727/43/6/065205.

[38] C. L. Enloe, T. E. McLaughlin, R. D. VanDyken, K. D. Kachner, E. J. Jumper, and T. C. Corke, "Mechanisms and responses of a single dielectric barrier plasma actuator: plasma morphology," *AIAA Journal,* vol. 42, no. 3, p. 589, 2004, doi: 10.2514/1.2305.
[39] N. Benard and E. Moreau, "Role of the electric waveform supplying a dielectric barrier discharge plasma actuator," *Applied Physics Letters,* vol. 100, no. 19, p. 193503, 2012.
[40] B.-R. Zheng, M. Xue, and C. Ge, "Dynamic evolution of vortex structures induced by tri-electrode plasma actuator," *Chinese physics B,* vol. 29, no. 2, p. 24704, 2020, doi: 10.1088/1674-1056/ab671f.
[41] K. Chen, X. Geng, Z. Shi, K. Cheng, and H. Cui, "Experimental investigation of influence of sliding discharge DBD plasma on low-speed boundary layer," *AIP advances,* vol. 10, no. 3, pp. 35108-035108-9, 2020, doi: 10.1063/1.5134848.
[42] E. Moreau, R. Sosa, and G. Artana, "Electric wind produced by surface plasma actuators: a new dielectric barrier discharge based on a three-electrode geometry," *Journal of Physics D: Applied Physics,* vol. 41, no. 11, p. 115204, 2008.
[43] A. Debien, N. Benard, and E. Moreau, "Electric wind produced by sliding discharges," *Proceeding of 2nd ISNPEDADM new electrical technologies for environment, Nouméa,* 2011.
[44] A. Tang, A. Aliseda, A. Mamishev, and I. Novosselov, "DC-Augmented Dielectric Barrier Discharge (DCA-DBD)," *arXiv preprint arXiv:2403.18064,* 2024.
[45] M. Forte, J. Jolibois, J. Pons, E. Moreau, G. Touchard, and M. Cazalens, "Optimization of a dielectric barrier discharge actuator by stationary and non-stationary measurements of the induced flow velocity: application to airflow control," *Experiments in Fluids,* vol. 43, no. 6, pp. 917-928, 2007, doi: 10.1007/s00348-007-0362-7.
[46] N. Benard, A. Mizuno, and E. Moreau, "A large-scale multiple dielectric barrier discharge actuator based on an innovative three-electrode design," *Journal of physics. D, Applied physics,* vol. 42, no. 23, p. 235204, 2009, doi: 10.1088/0022-3727/42/23/235204.
[47] S. Sato, H. Furukawa, A. Komuro, M. Takahashi, and N. Ohnishi, "Successively accelerated ionic wind with integrated dielectric-barrier-discharge plasma actuator for low-voltage operation," *Scientific reports,* vol. 9, no. 1, pp. 1-11, 2019.
[48] J.-C. Laurentie, J. Jolibois, and E. Moreau, "Surface dielectric barrier discharge: Effect of encapsulation of the grounded electrode on the electromechanical characteristics of the plasma actuator," *Journal of Electrostatics,* vol. 67, no. 2-3, pp. 93-98, 2009.
[49] D. M. Orlov, *Modelling and simulation of single dielectric barrier discharge plasma actuators*. 2006.
[50] T. C. Corke, M. L. Post, and D. M. Orlov, "Single dielectric barrier discharge plasma enhanced aerodynamics: physics, modeling and applications," *Experiments in Fluids,* vol. 46, no. 1, pp. 1-26, 2009.
[51] I. Biganzoli, R. Barni, and C. Riccardi, "Temporal evolution of a surface dielectric barrier discharge for different groups of plasma microdischarges," *Journal of physics. D, Applied physics,* vol. 46, no. 2, p. 025201, 2013, doi: 10.1088/0022-3727/46/2/025201.
[52] J. Kriegseis, B. Möller, S. Grundmann, and C. Tropea, "Capacitance and power consumption quantification of dielectric barrier discharge (DBD) plasma actuators," *Journal of Electrostatics,* vol. 69, no. 4, pp. 302-312, 2011.
[53] K. Kozlov, H. Wagner, R. Brandenburg, and P. Michel, "Spatio-temporally resolved spectroscopic diagnostics of the barrier discharge in air at atmospheric pressure," *Journal of Physics D: Applied Physics,* vol. 34, no. 21, p. 3164, 2001.
[54] V. V. Kovačević, B. P. Dojčinović, M. Jović, G. M. Roglić, B. M. Obradović, and M. M. Kuraica, "Measurement of reactive species generated by dielectric barrier discharge in direct contact with water in different atmospheres," *Journal of Physics D: Applied Physics,* vol. 50, no. 15, p. 155205, 2017.
[55] F. MacMillan, "Experiments on Pitot-tubes in shear flow," 1956.

[56] E. Moreau, C. Louste, and G. Touchard, "Electric wind induced by sliding discharge in air at atmospheric pressure," *Journal of Electrostatics,* vol. 66, no. 1-2, pp. 107-114, 2008.

[57] J. Deng, S. Matsuoka, A. Kumada, and K. Hidaka, "The influence of residual charge on surface discharge propagation," *Journal of Physics D: Applied Physics,* vol. 43, no. 49, p. 495203, 2010.

[58] C. L. Enloe, G. I. Font, T. E. McLaughlin, and D. M. Orlov, "Surface Potential and Longitudinal Electric Field Measurements in the Aerodynamic Plasma Actuator," *AIAA Journal,* vol. 46, no. 11, pp. 2730-2740, 2008, doi: 10.2514/1.33973.